\documentclass[10pt, conference, compsocconf]{IEEEtran}

\let\thanks\relax

\usepackage{cite}
\usepackage{amssymb}
\usepackage{amsmath}
\usepackage{amsthm}
\usepackage{url}
\usepackage[utf8]{inputenc}
\usepackage{graphicx}
\usepackage[hidelinks,bookmarks=false]{hyperref}
\usepackage{enumerate}
\usepackage{soul}
\usepackage{xcolor}
\usepackage{mathtools}
\usepackage{subcaption}
\usepackage{dblfloatfix}
\usepackage[font={small}]{caption}
\usepackage{microtype}
\usepackage{soul}
\usepackage[export]{adjustbox}
\usepackage{balance}
\usepackage{algorithmic}
\usepackage{algorithm}
\usepackage[pscoord]{eso-pic}
\newlength\myindent
\newcommand\bindent{%
  \begingroup
  \setlength{\itemindent}{\myindent}
  \addtolength{\algorithmicindent}{\myindent}
}
\newcommand\eindent{\endgroup}
\newcommand{\placetextbox}[3]{
  \setbox0=\hbox{#3}
  \AddToShipoutPictureFG*{
    \put(\LenToUnit{#1\paperwidth},\LenToUnit{#2\paperheight}){\vtop{{\null}\makebox[0pt][c]{#3}}}%
  }%
}%

\makeatletter
\def\blfootnote{\xdef\@thefnmark{}\@footnotetext}
\makeatother

\begin{document}

\title{Comprehensive Performance Analysis of\\Homomorphic Cryptosystems for\\Practical Data Processing}

\author{
  \IEEEauthorblockN{Vasily Sidorov\IEEEauthorrefmark{1}, ~ Ethan Wei Yi Fan\IEEEauthorrefmark{2}\textsuperscript{\#}, ~ Wee Keong Ng\IEEEauthorrefmark{3}}
  \IEEEauthorblockA{School of Computer Science and Engineering\\
    Nanyang Technological University\\
    Singapore\\
    Email: \IEEEauthorrefmark{1}\texttt{\href{mailto:vsidorov@ntu.edu.sg}{vsidorov@ntu.edu.sg}}, \IEEEauthorrefmark{2}\texttt{\href{mailto:ethanwei@ntu.edu.sg}{ethanwei@ntu.edu.sg}}, \IEEEauthorrefmark{3}\texttt{\href{mailto:awkng@ntu.edu.sg}{awkng@ntu.edu.sg}}}
}

\maketitle

\begin{abstract}
  Oblivious data processing has been an on and off topic for the last decade or so.
  It provides great opportunities for secure data management and processing, especially in the cloud.
  At the same time, modern computing resources seem to be affordable enough to allow for practical use of homomorphic cryptography.
  Yet, the availability of products that offer practical homomorphic data processing is extremely scarce.
  As part of a project aimed at developing a practical homomorphic data management platform, we have conducted an extensive study of homomorphic cryptosystems' performance, the results of which are presented in this work.
  For this work we chose the following five cryptosystems: fully homomorphic HElib and SEAL, somewhat fully homomorphic PyAono, and partially homomorphic Paillier and ElGamal.
  In the discussion of the aggregated results, we suggest that partially homomorphic cryptosystems could be used today in certain practical applications, whereas time has not yet come for the fully homomorphic ones.
\end{abstract}

\blfootnote{This work was supported by National Cybersecurity R\&D grant from National Research Foundation of Singapore.\\~\\
\textsuperscript{\#}\,At the time of publication at NUS--Singtel Cyber Security Research \& Development Laboratory, email:\,\href{mailto:dcsweiy@nus.edu.sg}{dcsweiy@nus.edu.sg}}


\IEEEpeerreviewmaketitle


\section{Introduction}
Oblivious data processing can solve a spectrum of security issues, mitigate a wide range of business risks, facilitate migration to the cloud for the companies that are still on the fence, and help with compliance in the view of new data protection regulations such as GDPR, PDPA or CCPA.
Cryptography is an obvious candidate on how to achieve obliviousness;
however, typically cryptosystems do have the capability for any meaningful manipulations with the encrypted data.

Homomorphic encryption is a type of encryption that is capable of carrying out specific types of computations directly on ciphertext and generates an encrypted result which, when decrypted, matches the result of operations would they have been performed on the plaintexts~\cite{yi2014homomorphic}.
Typically, homomorphic cryptosystems possess a homomorphic property with respect to one or several arithmetic operations, i.e., allowing for encrypted addition, multiplication, etc.
However, there also are some other types of homomorphic cryptosystems, e.g., Goldwasser-Micali probabilistic cryptosystem that allows for encrypted \texttt{XOR} (addition modulo 2)~\cite{goldwasser1984probabilistic}.

Strictly speaking, a homomorphic cryptosystem could be defined as follows.
Assume $e(\cdot)$ and $d(\cdot)$ to be the encryption and decryption functions, $v_1, \ldots, v_n$ to be plaintext values, and $f(\cdot)$ to be an $n$-ary plaintext operation. Then, if there exists an operation $g(\cdot)$ such that for any $v_1, \ldots, v_n$: $f(v_1, \ldots, v_n) = d(g(e(v_1),\ldots,e(v_n)))$, then the encryption scheme is said to have a property of homomorphism, i.e., it preserves operation~$f(\cdot)$.

For arithmetic homomorphic cryptosystems (ones that preserve arithmetic operations), we can distinguish two groups: \emph{partially} and \emph{fully} homomorphic cryptosystems.
Using partially homomorphic encryption, it is possible to perform only one operation on encrypted data---multiplication or addition---but not both~\cite{ogburn2013homomorphic}.
There are numerous ciphers capable of performing partially homomorphic encryption, but determining a feasible cipher for real world applications can be exceedingly challenging, primarily due to functional requirements and efficiency concerns.
A cryptosystem that is capable of performing both addition and multiplication over ciphertexts is called a fully homomorphic cryptosystem.

This work seeks to examine the practicality of homomorphic cryptography in the light of computing resources that are available today.
For this work we have chosen five implementations of homomorphic cryptosystems; namely, three fully homomorphic cryptosystems: Homomorphic Encryption Library (HElib), Simple Encrypted Arithmetic Library (SEAL), and an implementation of the \emph{somewhat}%
\footnote{Somewhat fully homomorphic scheme is a scheme that enables both addition and multiplication but exposes limits to the number of operations that can be executed.
Typically, this refers to a scheme that can perform many additions and a small amount (sometimes, just one) multiplications.
Some fully homomorphic schemes overcome this by executing a procedure called \emph{bootstrapping} --- an operation of ``refreshing'' the ciphertext.
Bootstrapping is usually a very expensive operation.
Unlike PyAono, both HElib and SEAL implement bootstrapping, which makes them fully homomorphic.}
homomorphic encryption scheme by Y.\ Aono \emph{et~al.}\ (PyAono); and our own implementations of Paillier (additive) and ElGamal (multiplicative) partially homomorphic cryptosystems.

This choice was influenced by several factors.
First, we were looking for solutions that reached a certain level of maturity and recognition in the research community, as an indication that both the underlying cryptographic scheme and the implementation can be considered thoroughly reviewed.
Cryptographic analysis of the schemes and analysis of implementation fidelity were not in the scope of this work.
In addition to that, we only considered implementations that had their license indicate that it is allowed to carry out research on them.
Furthermore, after extensive attempts we were still unable to get some of the available implementations to work due to compilation errors, lack of documentation, runtime errors, etc.
Those were eventually removed from the experiments.

The examination presented in this paper arose from our research on practicality of \emph{emulating} fully homomorphic computations using partially homomorphic cryptosystems and a trusted re-encryption agent equipped with a smart computation scheduler that enables re-encryptions between the partial schemes to be done while other portions of homomorphic computation are carried out.
In this research, our own implementations of Paillier and ElGamal are used as partially homomorphic cryptosystems in the emulation;
thus, they were also selected for the final list of contenders.
The work presented in this paper is a continuation of our earlier short analysis on the topic~\cite{towards}.

For each cipher, we obtained the source code and applied minor tweaks to measure precisely the computation time with the execution of each operation.
We additionally tested the performance of each cipher under a full spectrum of cryptosystem initialization parameters;
the data collected from these tests is presented in charts in the annex of the paper.

The rest of the paper is structured as follows:
Section~\ref{sec:related} reviews some of the previous works in a similar direction;
Section~\ref{sec:method} discusses details of the comparative testing that was carried out;
Section~\ref{sec:empirical} presents the findings and a short commentary;
Section~\ref{sec:conclusion} presents a concluding discussion of the results and indicates directions of future work;
performance charts are appended at the end of the paper.

\section{Related Work}\label{sec:related}
There is a multitude of homomorphic schemes: unpadded RSA~\cite{rivest1978method} and ElGamal~\cite{elgamal1985public} schemes are multiplicative homomorphic; Benaloh~\cite{benaloh1994dense}, Paillier~\cite{paillier1999public}, and Boneh-Goh-Nissim~\cite{boneh2005evaluating} schemes are additive homomorphic.
However, a fully homomorphic scheme, i.e., a scheme that supports both addition and multiplication, remained undiscovered for a long time and was referred to by some researchers as a ``holy grail of cryptography''~\cite{Micciancio:2010:FGC:1666420.1666445}.
A breakthrough in the area was made in 2009 when Dr.\ Craig Gentry has published his work on the first fully homomorphic encryption scheme~\cite{Gentry:2010:CAF:1666420.1666444}.

Most researchers agree that currently the Gentry's scheme is impractical for general-purpose computations due to its poor performance.
There have been multiple studies of its performance as well as attempts to improve it~\cite{SmartEfficiency,doi:10.1137/120868669,GentryEfficiency}, but it is still considered to be far from being practical.

Although several researchers have approached the issue of measuring the performance of homomorphic computations and comparing it to plaintext computations, there are not many empirical results in that direction.
Certain insights could be gathered from the work by Gentry \emph{et~al.}\ that covers implementation details of his scheme and how its behavior in practice~\cite{GentryEfficiency}.
An attempt to secure third-party applications by amending the binaries to work over homomorphically encrypted data was done by Tople \emph{et~al.}~\cite{Tople:2013:AEH:2508859.2516666}.
Performance analysis done in their project AutoCrypt could provide an approximate understanding of performance drop after switching from plaintext to homomorphic computations.
Thorough performance analysis of Gentry's scheme was done by Liu \emph{et~al.}~\cite{liu2010performance}.
However, the work omits to compare homomorphic computation performance to the plaintext alternative and focuses on the relation between the computation time and key length.

\section{Methodology}\label{sec:method}
Preliminary research was carried out to select the candidates for comparison.
Five solutions were identified based on existing use cases, code availability, license, and implementation stability.

\subsection{Homomorphic Ciphers}
Started in 2013, HElib is written in C++ and is still under active development by Shai Halevi (IBM), Victor Shoup (NYU, IBM), and others, and is one of the most comprehensive fully homomorphic encryption libraries available today.
It is built on top of the Brakerski-Gentry-Vaikuntanathan (BGV) lattice-based encryption scheme but delivers enhanced performance via the Smart-Vercauteren ciphertext packing and Gentry-Halevi-Smart homomorphic optimizations.
HElib is a prime candidate for research on appliances such as establishing cloud security, search engine query encryption, and spam filtering.
HElib is available on GitHub~\cite{HELib}.

SEAL project was started in 2015 but has changed its underlying cryptosystem to a variation of BFV cryptosystem~\cite{cryptoeprint:2012:144} with the release of v2.0 in 2016.
It was initiated in the Cryptography Research Group at Microsoft Research and is still undergoing active development.
Its engineering is mainly focused on providing a secure platform for cloud computing through fully homomorphic properties.
Implementation of SEAL, written in C++ and C\#, can be found on Microsoft's web page~\cite{SEALLib}.
In our experiments, we used v2.3.1 of the library.

PyAono is an implementation of an LWE-based homomorphic cryptosystem suggested by Y.\ Aono \emph{et~al.}~\cite{aono2015fast}.
PyAono is also available GitHub~\cite{PyAono}.
This repository contains a homomorphic library written in C++ and an interoperability layer for Python.
For the experiments, the homomorphic library was extracted and imported into a simplified C++ testing program, fully capable of measuring and logging computation time of each of its operations.

The Paillier cryptosystem was invented by Pascal Paillier in the year 1999~\cite{paillier1999public}.
It is a probabilistic asymmetric public-key encryption scheme with partially homomorphic features based on the problem of computing $n^\textrm{th}$ residue classes.
In its base form, the Paillier cryptosystem is only capable of homomorphic addition of ciphertexts and multiplication by a plaintext constant.
Paillier scheme happened to be especially popular in application to electronic voting systems, with the motive of creating a fast, secure, and just approach to collecting votes.
Our implementation of the Paillier cryptosystem is written in C\# and can be found on GitHub~\cite{PaillierExt}.

The ElGamal encryption system was developed and first presented by Taher ElGamal in 1985~\cite{elgamal1985public}.
Its cryptography is based on the computational Diffie-Hellman problem (CDH).
Similar to the Paillier cryptosystem, ElGamal encryption is a probabilistic asymmetric partially homomorphic public key encryption scheme.
ElGamal is a multiplicative partially homomorphic cryptosystem.
ElGamal scheme is mostly used for digital cryptographic signatures and more rarely for encryption; it is used in such popular systems as GNU Privacy Guard (GPG) and Pretty Good Privacy (PGP).
Our C\# implementation of ElGamal encryption can be found on GitHub~\cite{ElGamalExt}.

\subsection{Technical Details}
HELib, SEAL, and PyAono have support for three operations (addition, subtraction, multiplication, but not division) and support negative numbers out of the box.
However, a ``textbook'' version of Paillier cryptosystem only supports addition (no subtraction) of only positive integers.
Similarly, a textbook version of the ElGamal cryptosystem only supports multiplication of positive integers.
In our implementations we designed variations to the schemes to add support for inverse operations (subtraction and division) and for negative numbers.

In both cases, we implement a multi-step encoding.
As a first step, we convert the to-be-encrypted number $x$ to a rational (fixed-point decimal) number $\frac{x\times10^k}{10^k}$, where $k$ is some pre-defined constant ($k=12$ in our experiments), and store the number as a pair (numerator, denominator).
When encrypted, the numerator and denominator are encrypted separately:

\[e(x) = \left\{ e(x\times10^k), e(10^k) \right\}\]

This immediately enables division for ElGamal encryption: multiplication of $a = (n_1, d_1)$ and $b = (n_2, d_2)$ is $p = (n_1 \times n_2, d_1 \times d_2)$; division of $a$ by $b$ is $q = (n_1 \times d_2, d_1 \times n_2)$.
So, effectively, we reduced division to multiplication.
Paillier cryptosystem can operate on these encodings without any issues as well: as all numbers have the same denominator $10^k$, we can just add the numerators.

Clearly, this encoding also automatically enables fractional numbers with precision up to $k$ positions after the decimal point.
Upon decryption of the ciphertext, the number is converted back from a rational fraction by performing the final division of the decrypted numerator by the decrypted denominator.

The next step in the encoding process enables negative numbers.
For the encryption of negative numbers, we adopted the two's complement approach that is used in modern computers when dealing with signed integers, effectively transforming negative values to positive numbers before encryption.
With a predefined threshold $z = 2^i, i\in \mathbb{N}$, this scheme allows to encode numbers in the range from $[-z/2, z/2)$ by doing a 1-to-1 mapping of the interval $[-z/2, -1]$ to $[z/2, z-1]$.
Algorithm~\ref{alg:applytwo} illustrates this process.

\begin{algorithm}[t]
    \caption{En-/Decoding a number with two's complement}
    \label{alg:applytwo}
    \begin{algorithmic}
        \REQUIRE ~\\$n$: input parameter, the number to en-/decode\\$z$: max threshold, $z = 2^i, i \in \mathbb{N}$
        \STATE ~\\
        \textbf{function} Encode($n$)
        \bindent
            \IF{$n < 0$}
                \STATE $n \leftarrow n + z$
            \ENDIF
            \STATE \textbf{return} $n$
        \eindent
        \STATE ~\\
        \textbf{function} Decode($n$)
        \bindent
            \STATE $n \leftarrow n \mod z$
            \IF{$n > z/2$}
                \STATE $n \leftarrow n - z$
            \ENDIF
            \STATE \textbf{return} $n$
        \eindent
    \end{algorithmic}
\end{algorithm}

During decoding, the decrypted numbers undergo a modulus of $z$.
This is to reduce these numbers to their two's complement representation, which is a necessary step following any computation involving negative numbers in two's complement representation.
As an example, assume we have two 4-bit signed integers, $2$ and $-1$.
Their binary form in two's complement would be \texttt{0010} and \texttt{1111} respectively.
When we add $2$ and $-1$, the answer would be 1 in decimal form, but if we perform the same operation on their two's complement form, we will get $2 + 15 = 17$.
In this case, $z= 2^4 = 16$, and by evaluating $17 \mod z$ we obtain the correct answer of $1$.
If the result of modulus is larger than $z/2$, it will be converted back to its negative value by subtracting $z$ from it.

The prime limitation of this method is that the results of computation on ciphertexts must be within the range $[-z/2, z/2)$.
Hence, $z$ should be a value that is sufficiently large to store the expected results of the computation.
However, this behavior replicates the behavior of plaintext arithmetics implementation in modern computers, which is a useful property as the computation produces an expected result with under- and over-flows where they would occur in plaintext arithmetics as well.


\begin{figure*}[t]
    \centering
    \includegraphics[width=0.95\linewidth]{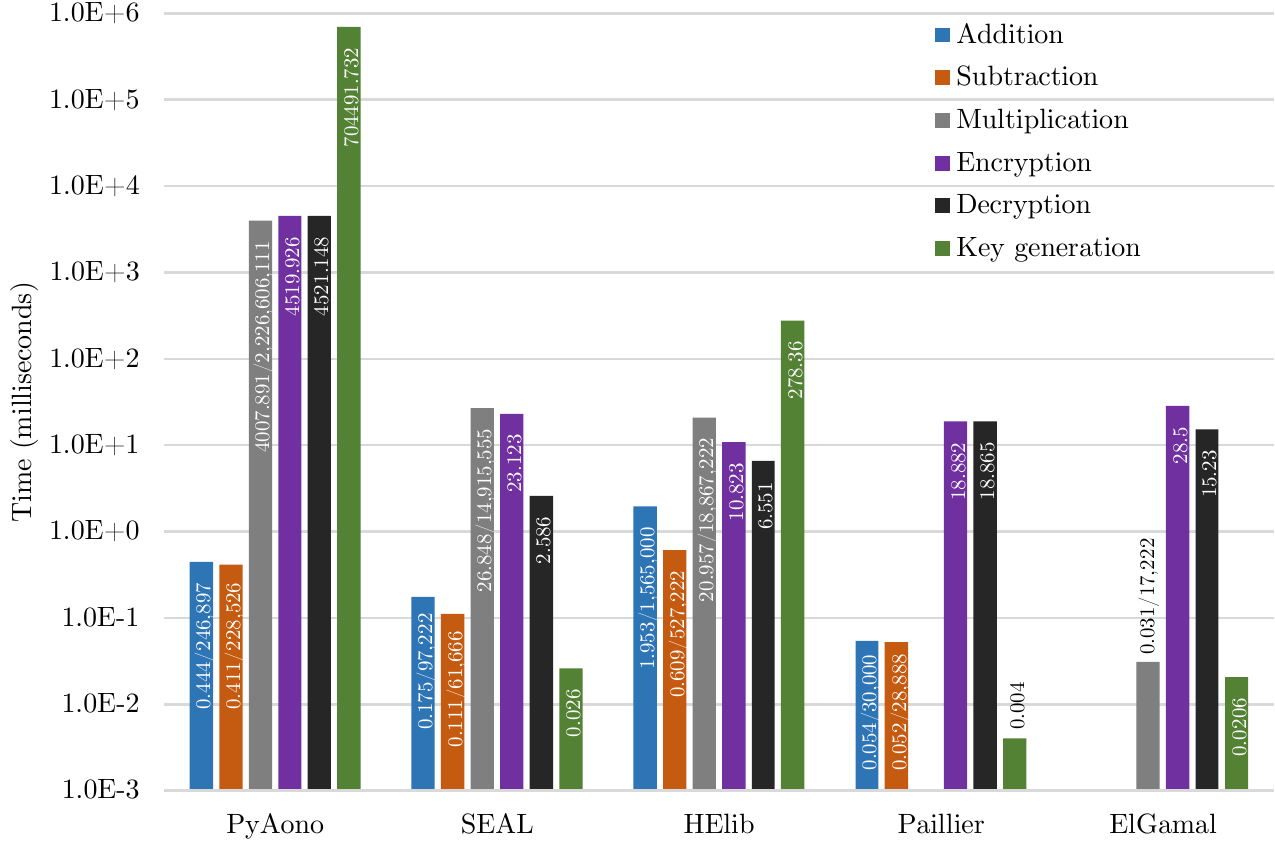}
    \caption{Overall times for all operations compared between each other with default parameters (1000 iterations).\\
             For operations of $+ - \times$, values are in form $t/r$, where $t$ is time in ms, and $r$ is the ratio of $t$ and the time execution of the same operation took over plaintexts.\\
             E.g., PyAono's addition is 246,897 times slower than plaintext addition.}
    \label{fig:17}
\end{figure*}

To ensure fairness and consistency in test results, we ran every  program that collected performance metrics on the same machine with an Intel Xeon E5-1650 v4 CPU with 6 cores running at 3.60GHz under Ubuntu 16.04.
We wrote a script to standardize the execution of all five ciphers and log the computation time for each run of the test program.
This allowed us to perform each test with the same general input, for instance, the number of iterations.

For each cipher, additional tests were conducted to determine the implications of each cipher's initialization parameters to its performance.
This was done in part to ensure that shifts in each parameter are impactful enough to have a separate record in the results and not too big to break the system.
This also helped to simplify the results of the experiment by identifying parameters with dependant values or values that were advised to be left unchanged.
Those were then removed from the scope of the experiment.

For every test, a single parameter of each cipher is changed, and each cipher is tested individually.
A control test was conducted using the default value of each parameter, to act as a point of reference in every test.
Data from all tests were recorded and charted, as demonstrated in sections below and in the Appendix.

\section{Empirical Results}\label{sec:empirical}
Due to the irregular range of data values collected, charts were plotted using a logarithmic scale.
Please note that differences in bar length are not linearly proportional to differences in computation time.

Tests were run on the three basic operations, namely addition, subtraction, and multiplication.
Division was left out due to SEAL, PyAono and Paillier not supporting division, whereas for ElGamal division is completely equivalent to multiplication.
We considered applying our number encoding scheme to support division in SEAL and PyAono, but eventually decided that the results are more informative if the implementations are tested in their ``vanilla'' form.

Results were obtained using operations on ciphertexts from all five ciphers, as well as on plaintext for scale.
All tests were conducted without bootstrapping (where applicable), as the strategy on deciding when to execute this computationally expensive procedure is non-trivial and varies from one use-case to another.
All three plain text operations ran at the same efficiency, which was on average $\approx1.8\times10^{-9}$ seconds per one execution, which is also $\approx10^4$ times as fast as the fastest homomorphic operation, which appeared to be multiplication by ElGamal.

All tests conducted were ran on 1000 pairs of 2-digit plaintext numbers.
Each test was repeated 5 times and the mean result was tabulated for each operation.
Aggregated results are presented in Figure~\ref{fig:17}.
Paillier possesses the fastest addition, subtraction, and key generation operations, ElGamal~--- the fastest multiplication operation, HElib demonstrates the fastest encryption, and SEAL has the fastest decryption.
PyAono was least efficient in four operations, namely multiplication, encryption, decryption, and key generation.
Its addition and subtraction operations are also relatively inefficient, topping only that of HElib.
During the research period, PyAono was declared deprecated, so it is unlikely to be optimized in the future.
We still include the empirical numbers for PyAono for reference, but we will not focus on it in discussing the results.

\section{Discussion \& Conclusion}\label{sec:conclusion}
Results clearly suggest that partially homomorphic cryptosystems are \emph{significantly} faster at homomorphic operations and relatively on par in en-/decryption times with HElib and SEAL.
Paillier and ElGamal are consistently about 20--30k times slower than plaintext.
At the same time, HElib is 500k times slower for subtraction, 1.6M times slower for addition, and almost 19M times slower for multiplication.
SEAL shows relatively acceptable performance on addition and subtraction: 60--100k times slower than plaintext; but 15M times slower for multiplication (not to forget that it doesn't support division).
PyAono takes the middle ground performing around 235k times slower than plaintext addition/subtraction; however, it absolutely breaks records in all other measurements: 2.2B~(sic!) times slower in multiplication, 4.5 seconds for en-/decryption and over 700 seconds~(sic!) for key generation.

Let us imagine that we use an emulation of fully homomorphic properties based on Paillier and ElGamal and a trusted re-encryption agent.
Re-encryption agent uses a serialized strategy of executing a computation: it performs all operations that are possible with current encryptions, then re-encrypts intermediate values into a different encryption scheme, and continues (see Algorithm~\ref{alg:execution} for an example).
Note that this strategy does not involve any parallelization; e.g., step~2 does not need to wait for the agent to finish encryption of $c$ and $d$, and the agent can start the decryption of $p$ in step~4 while step~3 is not yet complete.
We also do not take into account network latency and speed between the compute engine and the agent.

\begin{algorithm}[h]
    \caption{Execution plan for $(a + b) \times (c + d)$}
    \label{alg:execution}
    \begin{algorithmic}[1]
        \STATE \texttt{AGENT}: Paillier\_encrypt $a, b, c, d$
        \bindent
        \STATE \texttt{COMPUTE}: $p$ $\leftarrow$ Paillier\_add $a, b$
        \STATE \texttt{COMPUTE}: $q$ $\leftarrow$ Paillier\_add $c, d$
        \eindent
        \STATE \texttt{AGENT}: Paillier\_decrypt $p, q$
        \STATE \texttt{AGENT}: ElGamal\_encrypt $p, q$
        \bindent
        \STATE \texttt{COMPUTE}: $r$ $\leftarrow$ ElGamal\_multiply $p, q$
        \eindent
        \STATE \texttt{AGENT}: ElGamal\_decrypt $r$
    \end{algorithmic}
\end{algorithm}

Using the numbers that are presented in Figure~\ref{fig:17}, we can estimate the comparative efficiency of SEAL and HElib and the emulation described above on some sample tasks.
E.g., on a simple task of computing $r=(a \times b)+(c \times d)$, the emulation works with the same speed as SEAL and 1.46 times slower than HElib.

A slightly more complex task of computing a weighted sum of a large number of elements $\left(a_1 \times w_1 + \cdots + a_n \times w_n \right)$ is emulated at 0.9 the time of SEAL and is 1.35 times slower than HElib.
However, if we assume that the numbers were pre-encrypted (therefore eliminating the initial encryption), emulation finishes in 0.63 the time of SEAL and 0.75 the time of HElib.
Somewhere in between these two scenarios is such task as secure linear regression: the weights $w_i$ of the model are pre-encrypted, but the values $a_i$ that we are applying the model to are coming in online;
in this case, the emulation performs the computation in 0.81 the time of SEAL and 1.1 the time of HElib.

As we can see, on more complex tasks, the emulation can be on par or even better than fully homomorphic schemes.
It is important to notice that in reality, these theoretical numbers will be significantly corrected in favor of the emulation due to
\emph{1)}\;bootstrapping in fully homomorphic schemes, that \emph{will} happen on more complex tasks and is a \emph{very} expensive operation ($\approx$600\,ms bootstrapping a \textit{single} ciphertext in HElib as compared to 1--20\,ms for a homomorphic operation~\cite{ducas2015fhew});
\emph{2)}\;parallelization of work between the agent and the compute engine, which can significantly improve overall timings for the emulation.

It is also clear that due to the requirement in re-encryption under the emulation protocol, the en-/decryption efficiency of the partially homomorphic scheme becomes crucial for the overall performance.
We plan to continue our work on optimizing Paillier and ElGamal en-/decryption procedures and investigate other partially and fully homomorphic cryptosystems that could provide better overall performance.

\bibliographystyle{IEEEtran}
\bibliography{sources}

\balance


\clearpage

\placetextbox{0.865}{0.92}{\large Appendix}

\begin{figure*}[t]
    \centering

    \begin{subfigure}[b]{0.73\textwidth}
        \includegraphics[width=\textwidth]{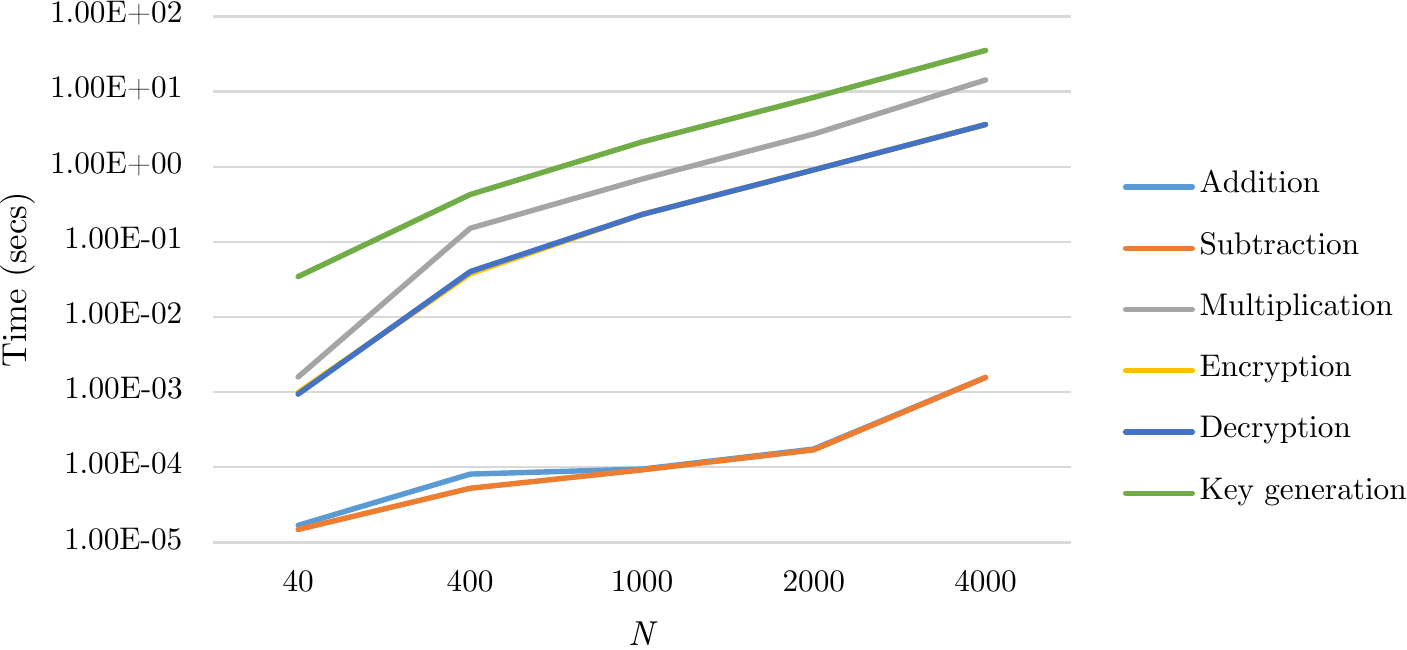}
        \caption{Varying parameter $N$, one of the two main security parameters\\~\\~}
        \label{fig:1}
    \end{subfigure}

    \begin{subfigure}[b]{0.475\textwidth}
        \includegraphics[width=\textwidth]{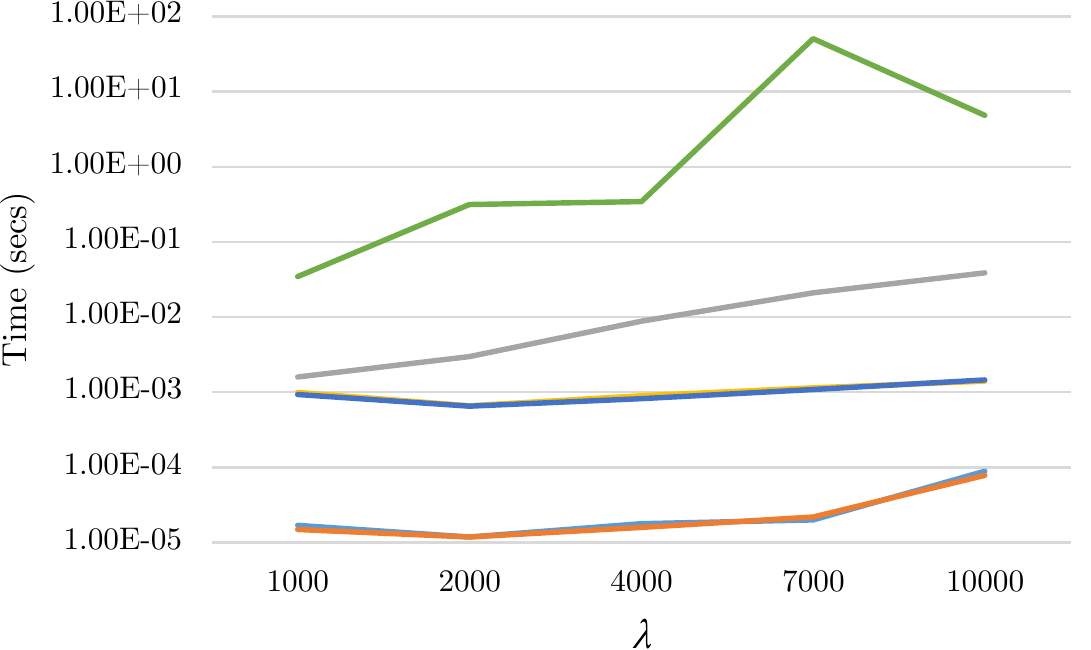}
        \caption{Varying parameter $\lambda$ (lambda), the second security parameter of the scheme\\~\\~}
        \label{fig:2}
    \end{subfigure}
    \qquad
    \begin{subfigure}[b]{0.475\textwidth}
        \includegraphics[width=\textwidth]{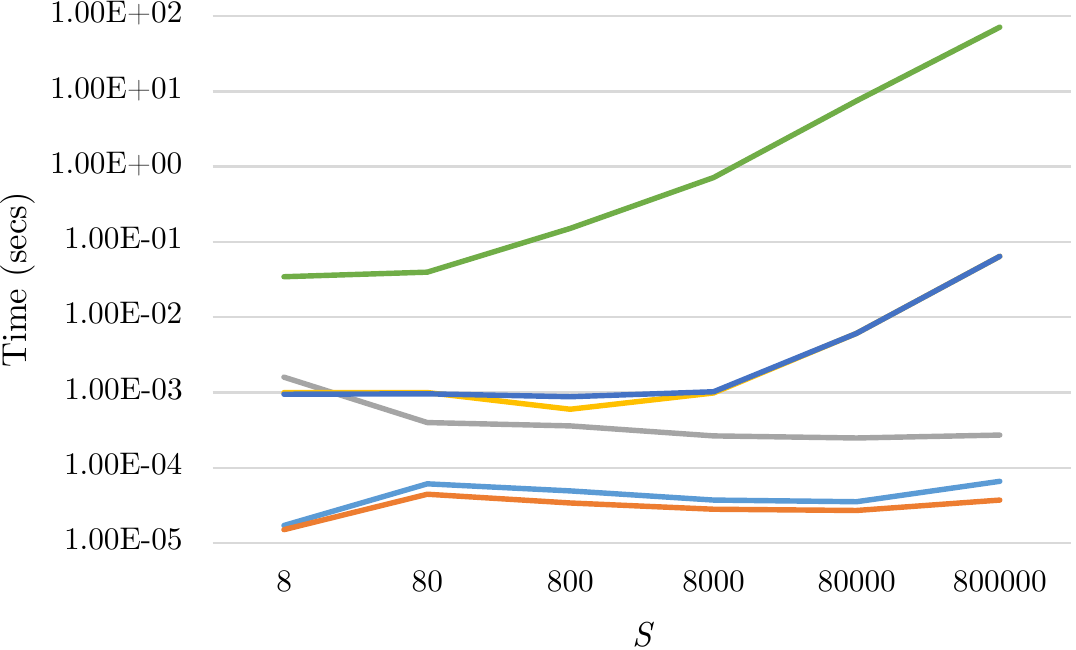}
        \caption{Varying a Gaussian noise parameter $S$, used in Knuth-Yao algorithm within the scheme\\~\\~}
        \label{fig:3}
    \end{subfigure}

    \begin{subfigure}[b]{0.475\textwidth}
        \includegraphics[width=\textwidth]{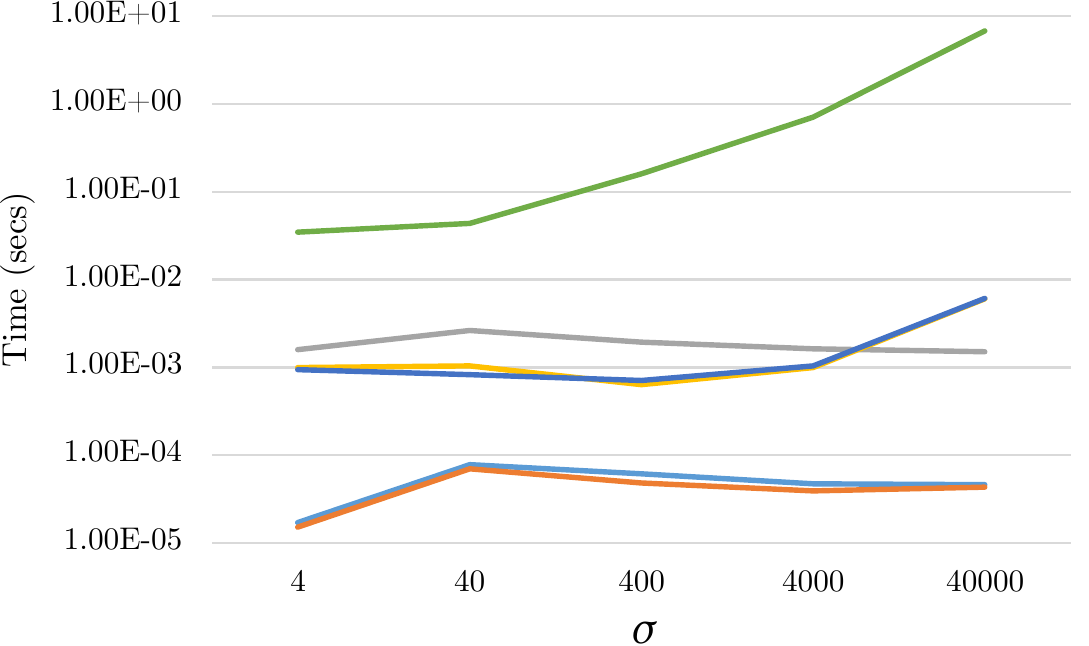}
        \caption{Varying the noise distribution parameter $\sigma$ (sigma)\\~\\~}
        \label{fig:4}
    \end{subfigure}
    \qquad
    \begin{subfigure}[b]{0.475\textwidth}
        \includegraphics[width=\textwidth]{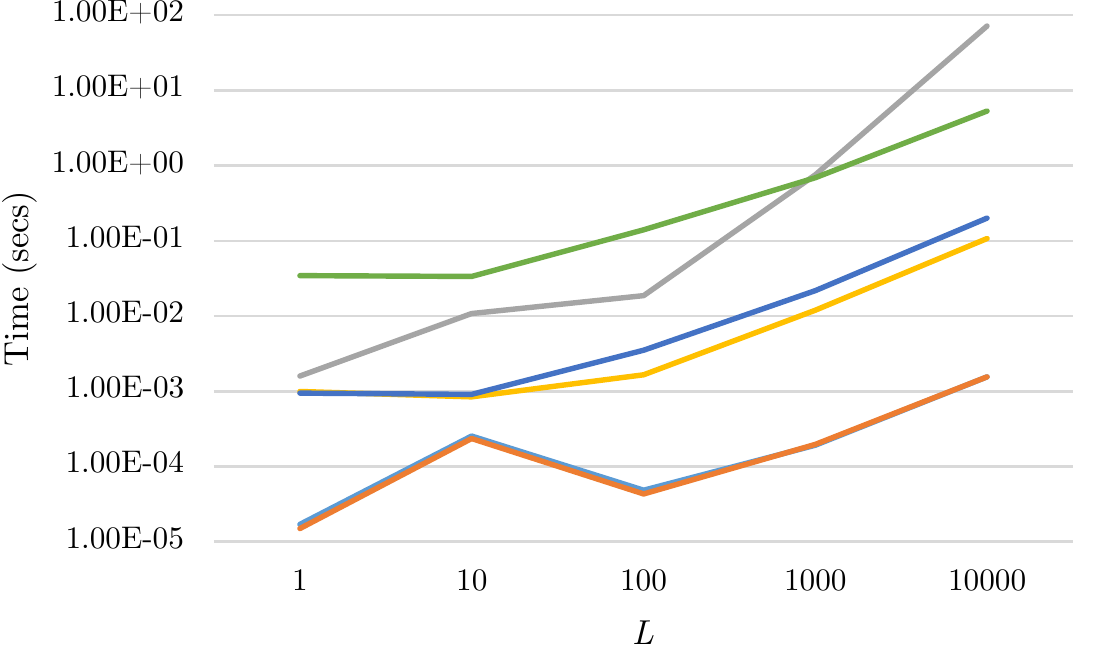}
        \caption{Varying the message length $L$\\~\\~}
        \label{fig:5}
    \end{subfigure}

    \caption{PyAono performance testing results}
\end{figure*}

\clearpage

\placetextbox{0.865}{0.92}{\large Appendix}

\begin{figure*}
    \centering

    \begin{subfigure}[b]{0.65\textwidth}
        \includegraphics[width=\textwidth]{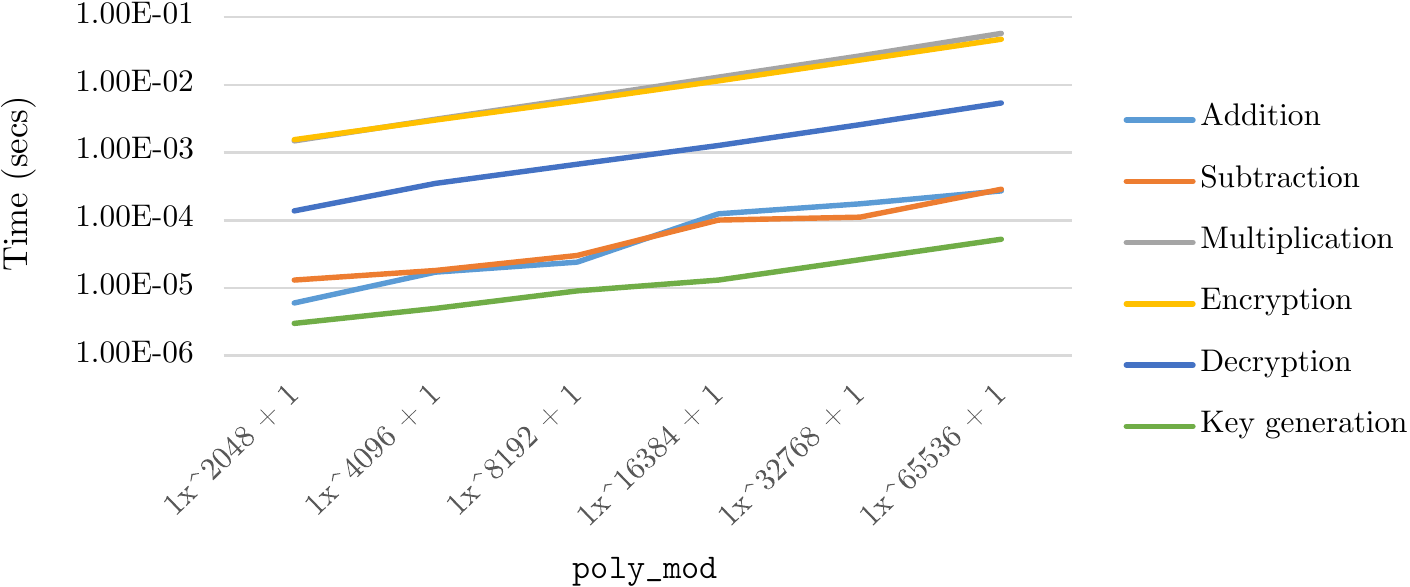}
        \caption{Varying the polynomial modulus that specififes the ring and affects security level\\~\\~}
        \label{fig:6}
    \end{subfigure}

    \begin{subfigure}[b]{0.475\textwidth}
        \includegraphics[width=\textwidth]{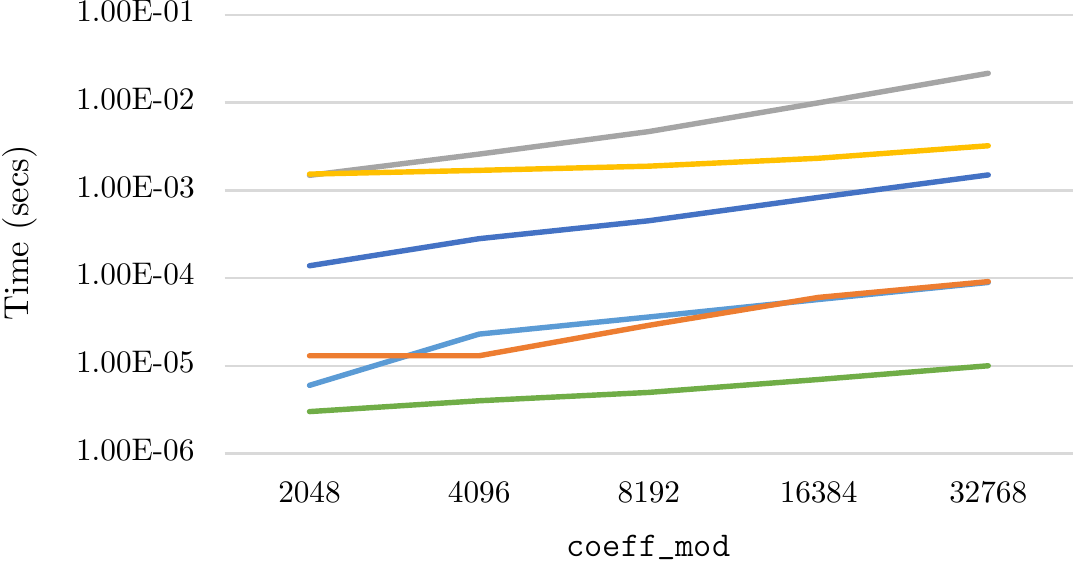}
        \caption{Varying the modulus in the cipher space}
        \label{fig:7}
    \end{subfigure}
    \qquad
    \begin{subfigure}[b]{0.475\textwidth}
        \includegraphics[width=\textwidth]{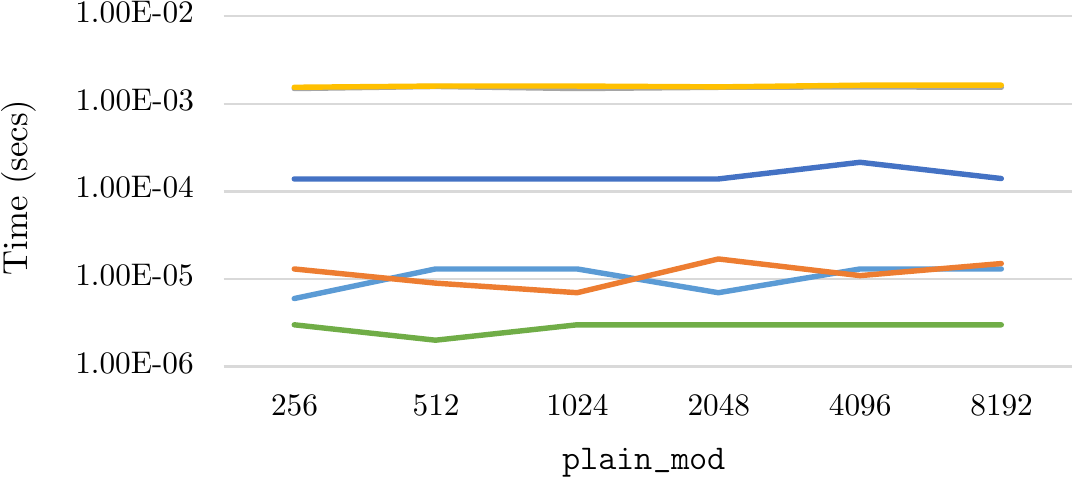}
        \caption{Varying the modulus in the plaintext space}
        \label{fig:8}
    \end{subfigure}

    \caption{SEAL performance testing results}
\end{figure*}

\begin{figure*}
    \centering
    \includegraphics[width=0.65\textwidth]{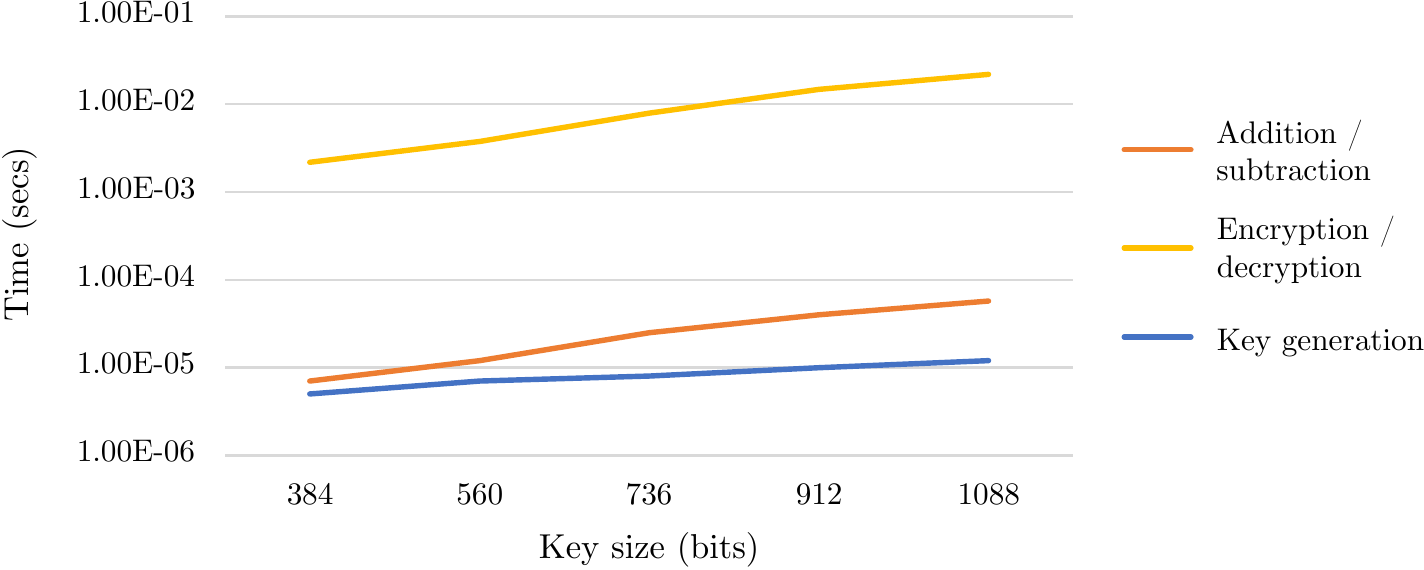}
    \caption{Paillier performance under varying the key size}
    \label{fig:9}
\end{figure*}

\begin{figure*}
    \centering
    \includegraphics[width=0.65\textwidth]{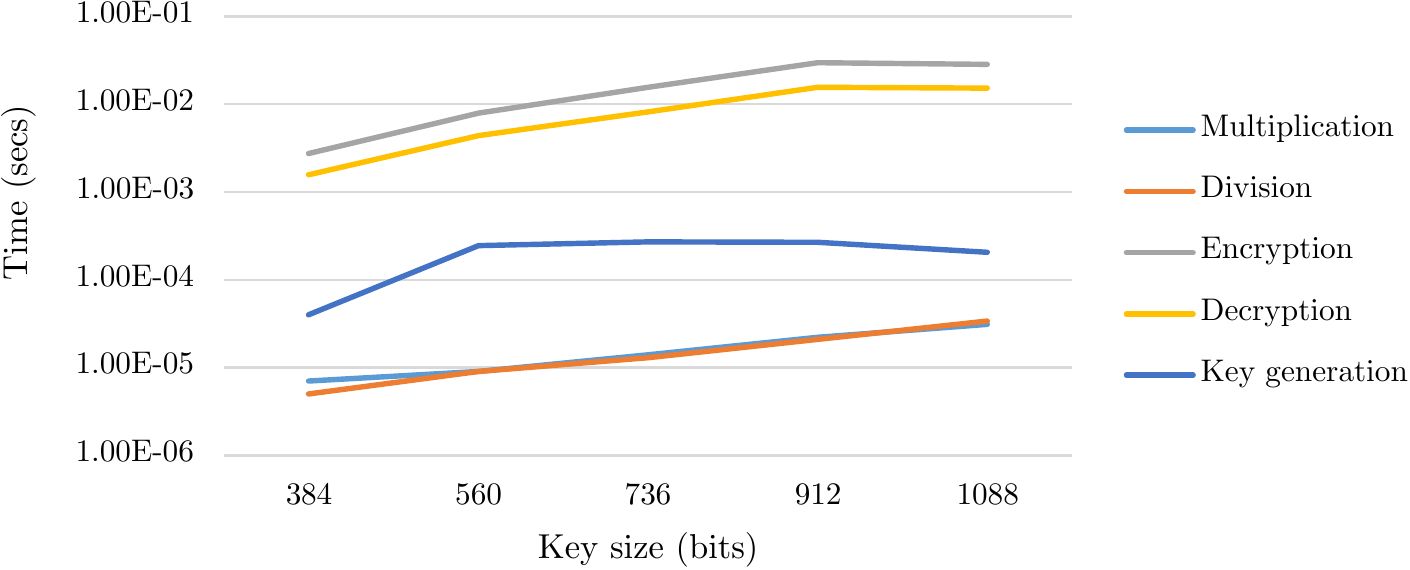}
    \caption{ElGamal performance under varying the key size}
    \label{fig:10}
\end{figure*}

\clearpage

\placetextbox{0.865}{0.92}{\large Appendix}

\begin{figure*}[t]
    \centering

    \begin{subfigure}[b]{0.17\textwidth}
        \includegraphics[width=\textwidth]{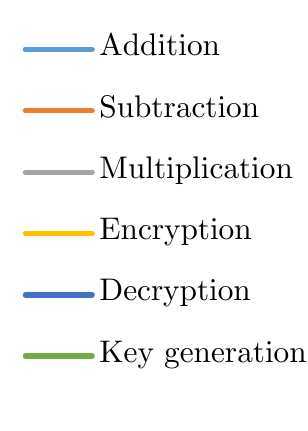}
        \label{fig:heliblegend}
    \end{subfigure}

    \begin{subfigure}[b]{0.475\textwidth}
        \includegraphics[width=\textwidth]{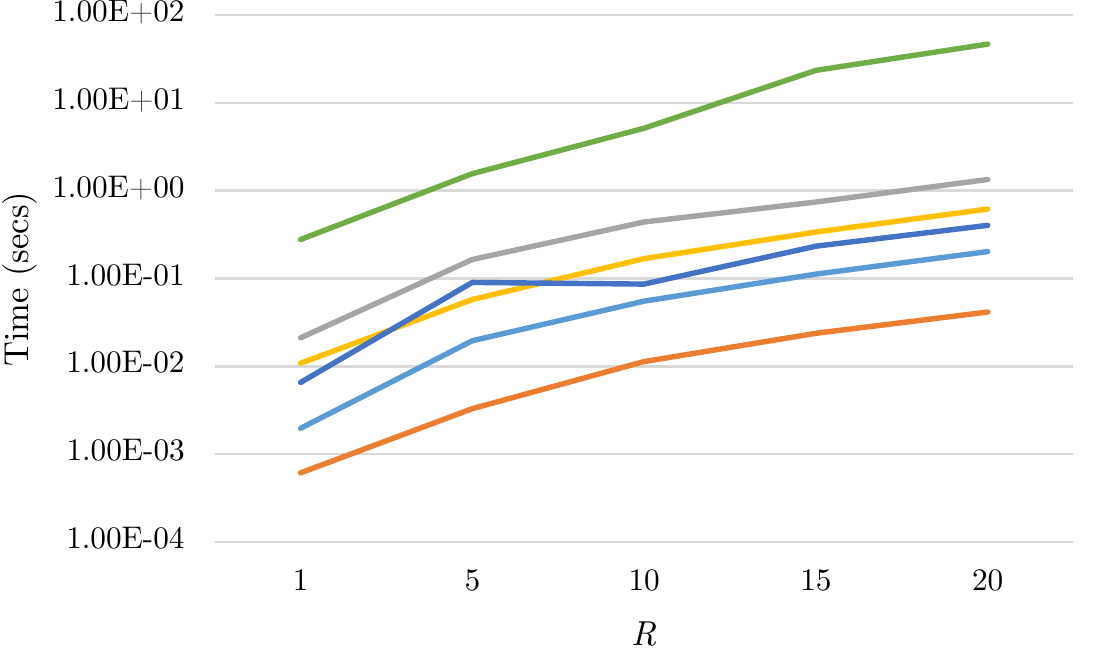}
        \caption{Varying the number of primes in the modulus chain $R$\\~\\~}
        \label{fig:11}
    \end{subfigure}
    \qquad
    \begin{subfigure}[b]{0.475\textwidth}
        \includegraphics[width=\textwidth]{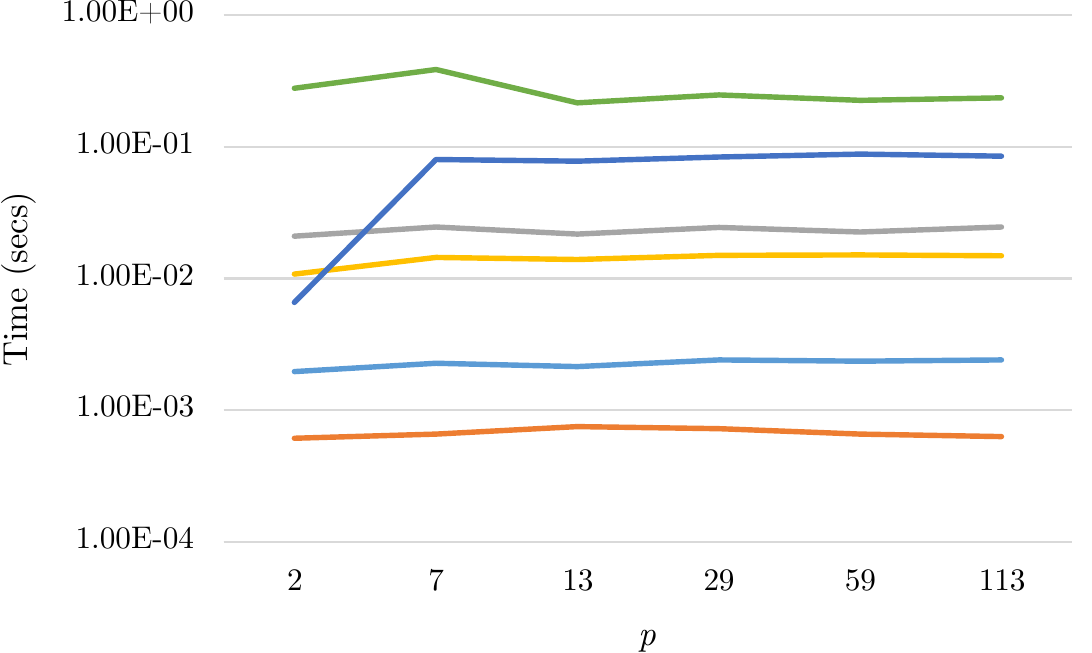}
        \caption{Varying the plaintext base $p$\\~\\~}
        \label{fig:12}
    \end{subfigure}

    \begin{subfigure}[b]{0.475\textwidth}
        \includegraphics[width=\textwidth]{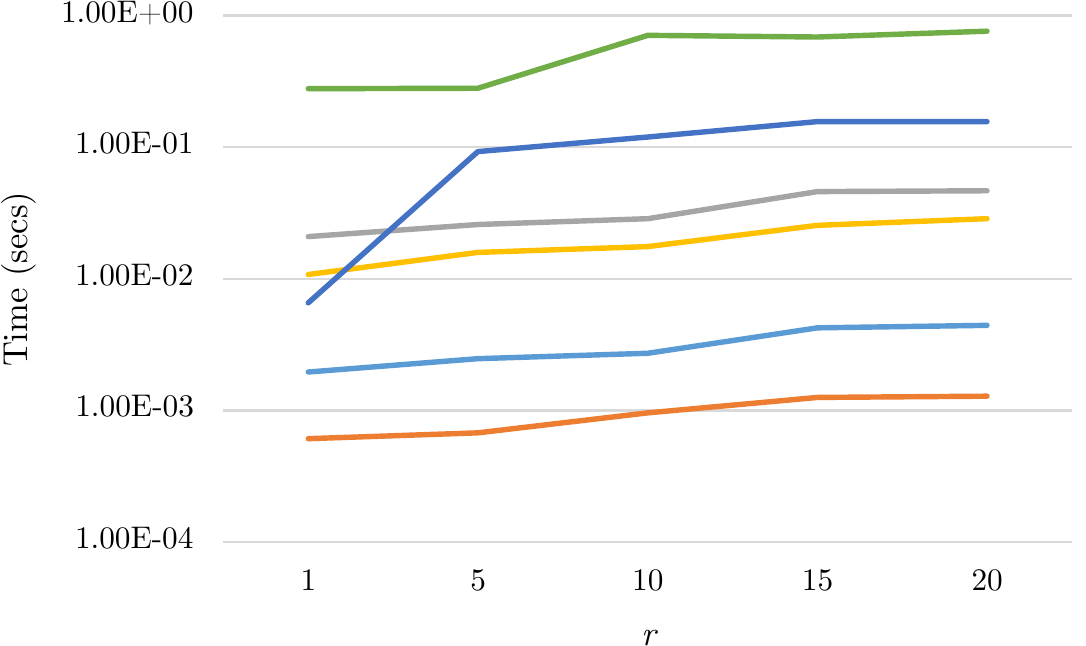}
        \caption{Varying the bootstrapping parameter (lifting) $r$\\~\\~}
        \label{fig:13}    
    \end{subfigure}
    \qquad
    \begin{subfigure}[b]{0.475\textwidth}
        \includegraphics[width=\textwidth]{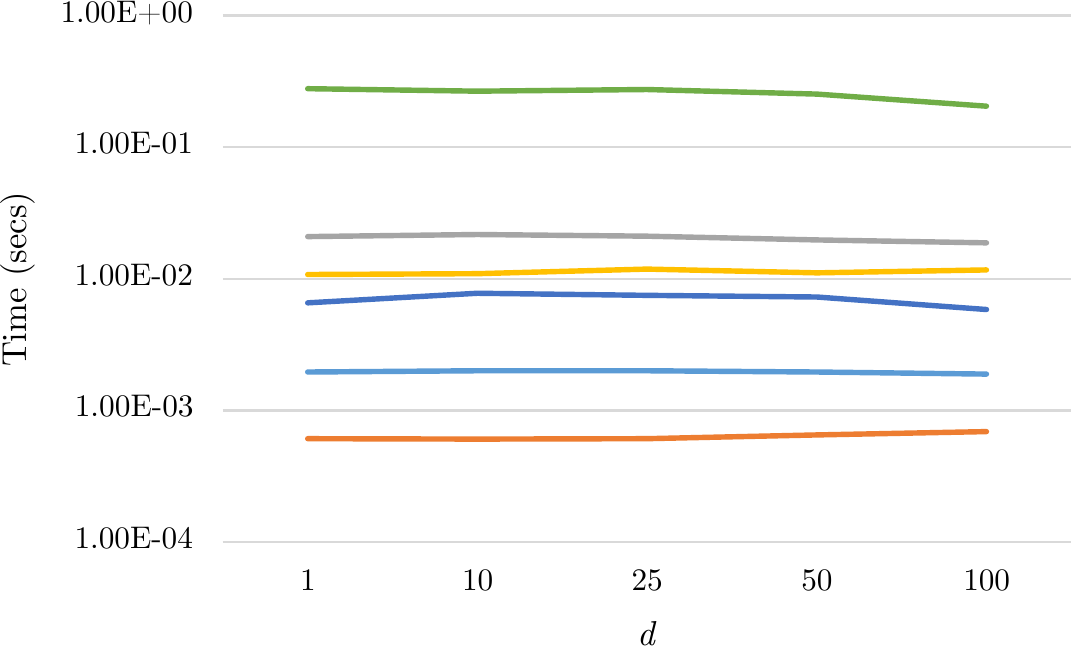}
        \caption{Varying the degree of field extension $d$\\~\\~}
        \label{fig:14}    
    \end{subfigure}

    \begin{subfigure}[b]{0.475\textwidth}
        \includegraphics[width=\textwidth]{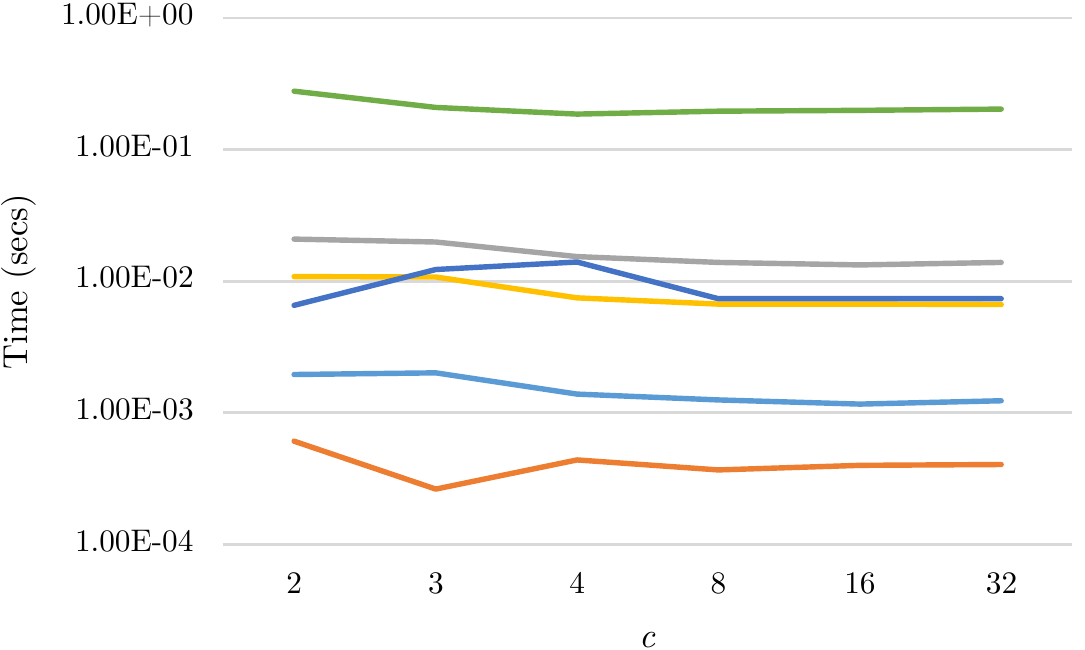}
        \caption{Varying the number of columns in the key-switching matrices $c$}
        \label{fig:15}    
    \end{subfigure}
    \qquad
    \begin{subfigure}[b]{0.475\textwidth}
        \includegraphics[width=\textwidth]{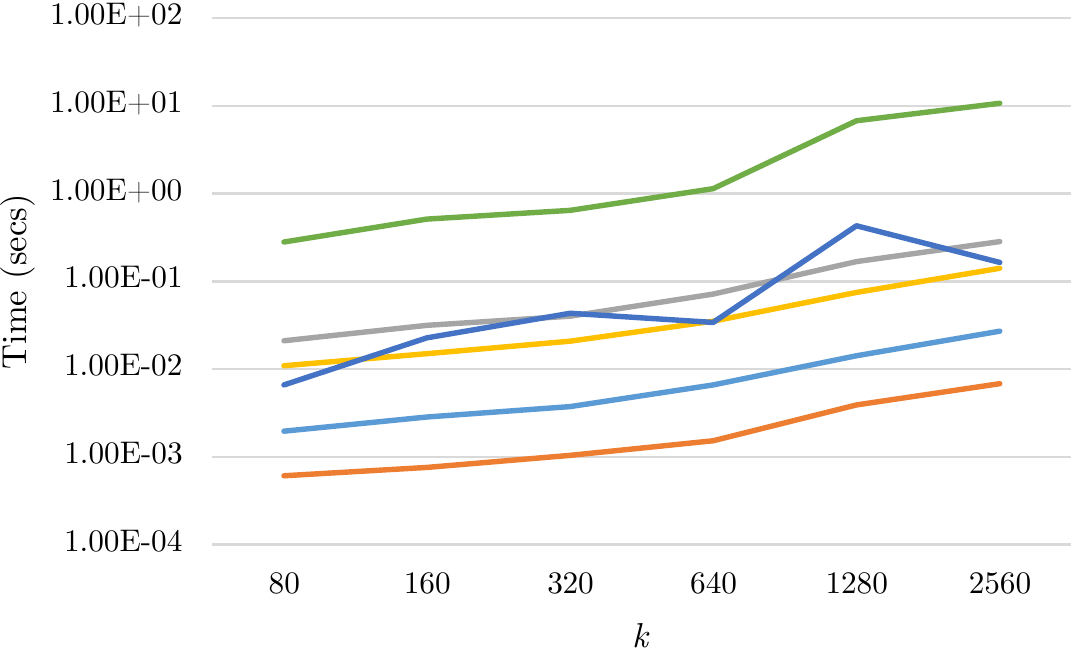}
        \caption{Varying the main security parameter $k$}
        \label{fig:16}
    \end{subfigure}
        
    \caption{HElib performance testing results}
\end{figure*}

\end{document}